\documentclass[10pt]{article}  
\usepackage[T1]{fontenc}
\usepackage{tgpagella}
\usepackage{authblk}
\usepackage[top=4cm, bottom=4cm, left=3cm, right=2.5cm]{geometry}
\usepackage{footmisc}
    
\usepackage{float}
\usepackage{graphicx}
\usepackage{setspace}
\usepackage{csquotes}
\usepackage{hyperref}
\usepackage{xcolor}
\usepackage{amsfonts}
\usepackage{amsmath}
\usepackage{tikz}

\usepackage[backend=bibtex,style=windycity,sorting=nyt,autocite=inline]{biblatex}
\newtheorem{definition}{Definition}
\newtheorem{example}{Example}

\title{Reflexive Measurement}
\author{James Michelson}
\affil{Department of Philosophy, Carnegie Mellon University}
\date{\today}

\begin{document}

\maketitle

\begin{abstract}
This essay is the first systematic account of causal relationships between measurement instruments and the data they elicit in the social sciences. This problem of reflexive measurement is pervasive and profoundly affects social scientific inquiry. I argue that, when confronted by the problem of reflexive measurement, scientific knowledge of the social world is not possible without a  model of the causal effect of our measurement instruments. 
\end{abstract}



\section{Introduction}

The idea of a ``self-fulfilling prophecy'' is an ancient one, going back at least as far as the story of Oedipus. In its more modern, philosophical guise philosophers of science have called the idea \textit{reflexive prediction}. A contemporary example of this phenomenon is bank runs: announcing an impending bank run may incite one. Contemporary philosophers of  science have stressed the challenge reflexive prediction poses for theory development and testing in the social sciences (see \autocite{kopec2011, lowe2018}). Less explored, however, is the idea of \textit{reflexive measurement}. A reflexive measurement is one that causally affects what the social scientist is trying to measure. Thus, the act of measuring some quantity about people (their beliefs or opinions, patterns of behavior, preferences, etc) may induce them to alter their behavior and thus ultimately produce a different measurement. This essay offers the first systematic account of this idea.

The importance and ubiquity of measurement issues in the social sciences which broadly fall under reflexive measurement bears emphasising. Examples can be found in all branches of social scientific inquiry. The problem of asking leading questions in political surveys---often called \textit{priming}---undermines the conclusions drawn from survey data \autocite{groves2011}. Psychological studies face problems of \textit{demand characterization} where participants try to infer what the study designer wants and then exhibit that behaviour \autocite{orne1962}. More broadly, the \textit{Hawthorne effect}---the change in individuals' behavior from merely being observed---in social scientific research falls under the idea of reflexive measurement and this idea is even found in adages like `Goodhart's law'\footnote{Goodhart's law is often stated as ``When a measure becomes a target, it ceases to be a good measure''.}. In all these cases, the act of measurement itself changes something about the social world (e.g., people's willingness to lie, likelihood participation in a study, exhibited behaviour, goals or incentives, etc).

I defend the claim that social scientists need scientific theories of their measurement instruments. In many settings we explore below, I believe social scientists implicitly do this. Too often, however, the automatic response of a social scientist confronted with measurement issues is to reach for statistical correction (e.g., de-biasing systematic (measurement) error). Without a deeper, philosophically-motivated understanding of measurement concerns in social scientific research no purely statistical technique can ultimately correct for problems of reflexive measurement. Recognition of the problem of reflexive measurement forces social scientists to consider measurement concerns in an entirely new light.

Although the concept of reflexive measurement as developed here is new, the idea that acts of measurement casually effect social phenomena under investigation is neither novel in the philosophy of social science nor the social sciences themselves. The view of `measurement as intervention' advocated here has origins in foundational works of philosophy of science:

\begin{quote}
  ``Scientific investigation is typically carried on in a noisy environment; an environment in which the data we confront reflect the operation of many different causal factors, a number of which are due to the local, idiosyncratic features of the instruments we employ (including our senses) or the particular background situation in which we find ourselves.'' \autocite[p398]{woodward1989}
\end{quote}

\noindent With regards to the origin of reflexive prediction in the social sciences, the idea is commonly traced back to the sociologist Robert K. Merton \autocite{merton1948}. In the mid-Twentieth Karl Popper and Ernst Nagel explicitly addressed Merton's problem of \textit{self-fulfilling prophecies} and reflexive predictions were an active topic of debate in philosophy of science in the 1960s and 1970s (see \autocite[p2-3]{lowe2018} for discussion). Central to these debates were the ideas of \textit{theory} and \textit{prediction}. Measurement concerns did not feature. Thus, I draw from the work of contemporary philosophers of economics to ground my understanding of measurement instruments in the social sciences. These accounts emphasize the diversity of measurement instruments instruments used in the social sciences \autocite{morgan2001}.

The issues introduced here are explored in more detail below. In section \ref{sec_problem} I define reflexive measurement and give a wide array of examples to illustrate how it affects social scientific research. Particular attention is given to the idea of a \textit{data generating process} which underpins statistical inference. Additionally, we provide an extended discussion of how reflexive measurement relates to the statistical concept of measurement error. In section \ref{sec_social_science} I focus more narrowly on measurement instruments in the social sciences and examine how social scientists might develop theories of their instruments. I contrast this with the naive use of robust statistical methods to tackle measurement issues in social science research. In section \ref{sec_conc}, I conclude with observations about the state of contemporary social scientific research and reflect on some of my concrete proposals. I defer a discussion of the problem of reflexive measurement in qualitative social scientific research until the conclusion.

\section{The Problem of Reflexive Measurement}\label{sec_problem}

It is helpful to fix certain background ideas before discussing reflexive measurement. In particular, the distinction between \textit{data} and \textit{phenomena} is especially important on the account proposed here. Phenomena are ``relatively stable and general features of the world which are potential objects of explanation and prediction by general theory'', whereas data ``by contrast, play the role of evidence for for claims about phenomena'' \autocite[p393-4]{woodward1989}. What matters in any scientific description or analysis of a phenomena is that ``the data should be \textit{reliable evidence} for the phenomena in question'' \autocite[p398, emphasis original]{woodward1989}. To make things concrete, an example of a phenomena which a social scientist would investigate is that of high school completion rates.

\begin{example}[High School Completion Rates]\label{ex_high_school}
The phenomenon under investigation by a social scientist is the lack of the completion of high school in the United States. Data might come from a survey question which asks respondents if they completed high school. Alternatively, the data could be aggregated official school attendance records.
\end{example}

\noindent No social scientist would expect that the high school completion rate obtained from survey data and that obtained by official attendance records should match. In fact, they might differ wildly. (Official records are often more accurate than self-reported data from surveys.) If these data sources differed, it would likely be an indication that self-reported dropout rates from surveys may not be ``reliable evidence'' (above) for the phenomena. 

Measurements of a phenomena produce data\footnote{For an extended account of measurement quantities, indications, and outcome (albeit with a particular emphasis on the natural sciences) see \autocite{tal2019}.}. In the social sciences, the act of measurement often interferes with what is being measured in some way, potentially undermining the role of data as ``reliable evidence'' for the phenomenon. This observation motivates the concept of reflexive measurement, here defined as:

\begin{definition}[Reflexive Measurement]
An act of measurement is reflexive if and only if it can causally effect either the phenomenon under investigation or an associated data generating process.
\end{definition}

\noindent We define a data generating process below, as well as outline how a measurement might causally affect it. To make vivid how an act of measurement may cause the phenomena under investigation to change, consider the following example.




\begin{example}[New York Sociologist]\label{NYC}
A sociologist in New York city wishes to ascertain whether people around the world have a favourable opinion of New York city. The sociologist flies his study participants from wherever they live to New York and conducts an in-person interview.
\end{example}

For many people in the world a flight to New York city may be a life-changing experience. Whatever their initial opinions of New York city, flying across the world to submit to an interview may radically change their opinions. Notice that this example is implausible as mode of social science research because of our almost intuitive understanding of the social world: individuals' cultural attitudes may conflict with those of the United States, their expectations may color their subsequent experience, openness to new experiences can very drastically between people can cultures, etc. These intuitions are enshrined in textbooks for social science research best practices (see for example, \autocite{groves2011, king2021}) and should be considered proto-theoretical social scientific knowledge of our social world. In this case our intuitions guide our judgement that this interview is unlikely to provide ``reliable evidence'' for the phenomenon of favorable attitudes towards New York City.

Another example which further illustrates the idea of measurement as intervention also motivates the incorporation of the `act of measurement' in the definition of reflexive measurement. Persons or institutions conducting the research may themselves affect the reliability of the data. By defining reflexive measurement at the level of an act of measurement (and not at the level of, say, measurement instruments) we can account for the following important example involving well-known institutions.


\begin{example}[ICE survey]\label{ICE}
The United States (US) Immigration and Customs Enforcement (ICE) wants to ascertain where undocumented immigrants reside. ICE conducts a phone survey across the US where the person answering the phone is asked whether they are an undocumented immigrant. The survey is conducted such that before the respondent is asked a question they are told ``This is a survey on behalf of The United States Immigration and Customs Enforcement''.
\end{example}

Are we to believe the results of this survey? ICE is one of the US government agencies directly responsible for deporting undocumented immigrants in the United States. Any survey or measurement they take concerning the number or location of undocumented immigrants is not likely to be reliable evidence of the phenomenon of undocumented immigration. This is a particular feature of the institution (i.e., ICE) asking about undocumented immigrants. It is in an undocumented immigrant's best interest to withhold this information from ICE. (In contrast, it might not be in an undocumented immigrant's best interest to withhold information if ICE were to, say, conduct a survey on popular ice cream flavors\footnote{Many groups of people may believe that answering any survey by ICE increases their risk of an undesirable outcome. This \textit{differential non-response} will affect the sample composition of ICE's survey. The important point is that the specific dynamics exhibited by the survey instrument in example \ref{ICE} no longer apply.}.) The person or institution conducting social scientific research may influence the data collected just as much as the choice of measuring instrument. 

Example \ref{ICE} reveals another feature of reflexive measurement which requires further scrutiny. The act of measurement did not change the number of undocumented workers in the US. The underlying phenomenon of interest is unchanged by the measurement. However, the data collected are no longer reliable evidence for the phenomena in question. To make sense of this we introduce the concept of a \textit{data generating process}\footnote{This is often called a \textit{data model} by statisticians (see \autocite[\S 4.1]{breiman2001} for discussion of proliferation of data models in statistical research).}. With this concept we can begin to make sense of the idea that an act of measurement causally affects the data collected without affecting the underlying phenomena under investigation.

Recall that measurement instruments produce data and this data is (hopefully) reliable evidence for the phenomenon. In example \ref{ex_high_school} I noted that data for the phenomenon of high school drop out rates could come in the form of survey responses or official school attendance records. Data could come in many other forms. However, in order to use the data for evaluating scientific theories, additional assumptions about the data need to be introduced. These assumptions might be statistical (i.e., the data follow some known distribution) or they might capture the process or system that generated the data. These assumptions are idealizations concerning how the phenomenon under investigation produced the observed data (i.e., the raw numerical values). It is these assumptions that are the data generating process. There are two ways that social scientists might use the term `data generating process':

\begin{enumerate}
    \item In it's \textit{statistical} sense, it is most often used as a linguistic shortcut to introduce a statistical description of the data at the level of random variables. For example, ``the random variables $X_i, \dots, X_n$ corresponding to the data from participants in our study are generated by a process best described by a Gaussian distribution with unknown mean $\mu$ and variance $\sigma$'' (often written ``$X_1, \dots, X_n \sim \mathcal{N}(\mu, \sigma)$'', where ``$\sim$'' reads `distributed as'.)

    \item Social scientists sometimes use the phrase in \textit{mechanistic} sense often devoid of statistical meaning. This might capture the idea that the data are generated by a social system best described by a number of differential equations or a multi-agent simulation model.

\end{enumerate}

\begin{figure}[t]
    \begin{center}
    \includegraphics[width=0.49\textwidth]{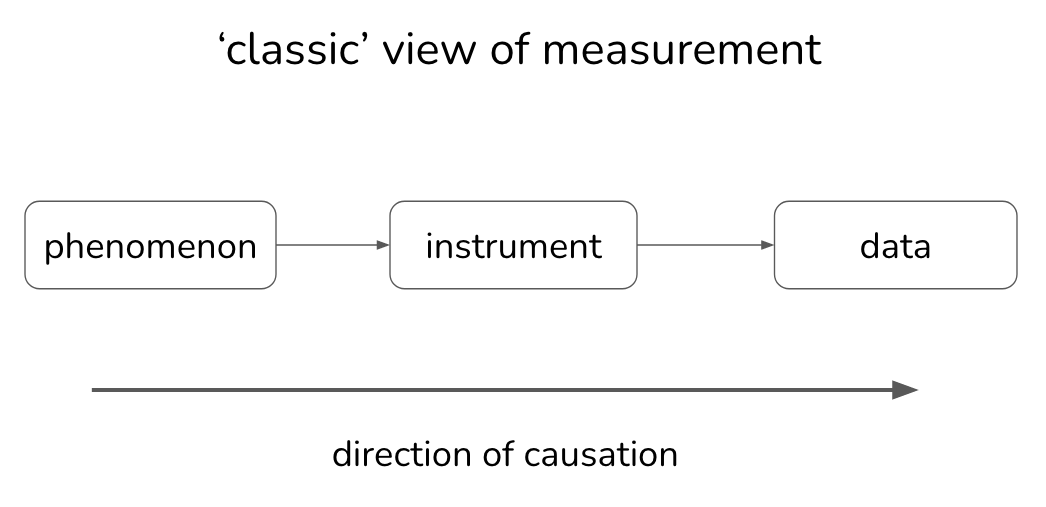}
    \includegraphics[width=0.49\textwidth]{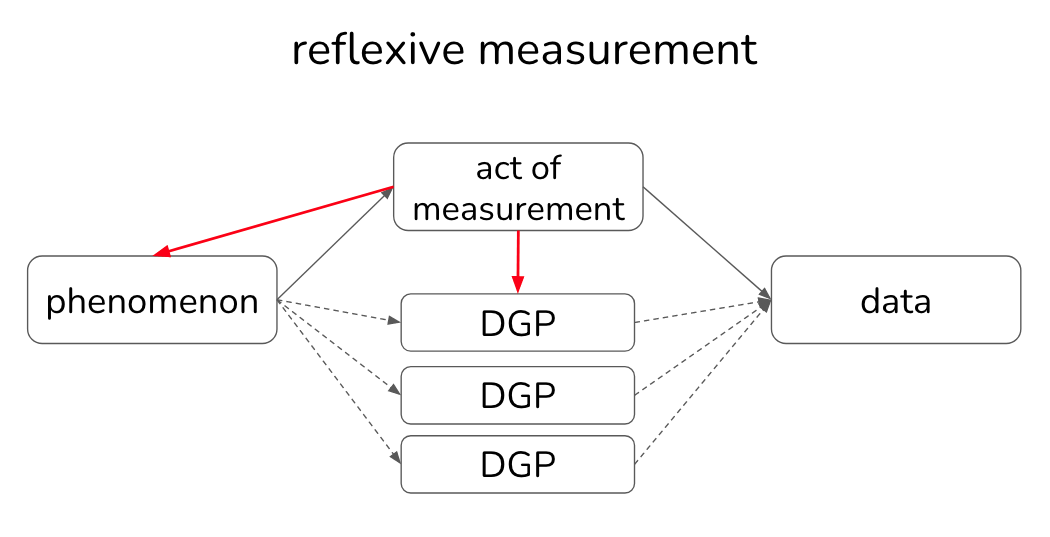} \\
    \end{center}
    \doublespacing{{\sc Figure 1\label{fig1}:} The figure on the left shows the classic view of measurement in the natural sciences \autocite{tal2019}. The figure on the right shows the effect of reflexive measurement on phenomenon under investigation and the data generating process (DGP) with red arrows.}
\end{figure}

A single phenomenon can be analyzed by multiple data sets. A single data set can be described by multiple data generating processes (of either kind above). The phrase is part technical jargon, part metaphysical claim. It is an idealization of a phenomenon which facilitates its (often statistical) analysis. A data generating process is causally upstream from the data but downstream from the phenomenon under investigation (see \ref{fig1}). 

Returning to the ICE survey in example \ref{ICE}, we can see that reflexive measurement which does not causally effect the phenomenon can still causally effect the data. This has been recognised for decades. Writing about the role of measurement instruments in economics, Mary Morgan writes: 

\begin{quote}     
    ``The ways in which the economic body is investigated and data are collected, categorized, analyzed, reduced, and reassembled amount to a set of experimental interventions---not in the economic process itself, but rather in the information collected from that process.`` \autocite[p237]{morgan2001}
\end{quote}

\noindent I believe it is exactly this casual effect of reflexive measurement on the data generating process which best captures Morgan's ``interventions''.

By way of contrast, consider the statistical notion of \textit{measurement error}. This framework can capture casual effects of the instrument on the data generating process at the level of \textit{systematic error}. The systematic component of measurement error always occurs, with the same value, when the instrument is used in the same way in the same case. (See, \autocite{tal2019} for philosophical discussion of types of measurement error.) Thus, for example, we might say a the systematic component of measurement error for a poorly-worded survey question on political attitudes occurs when the responses are, for example, `X\% more left/right-leaning'.

\begin{example}[De-Biasing Measurement Error]\label{ex_measure}
Consider the case of a survey question asking about presidential approval in the US, which was answered by $n$ respondents. The data (i.e., the random variables $X_1, \dots, X_n$) collected from this survey follow a data generating process given by: $X_1, \dots, X_n \sim \mathcal{N}(\mu,\sigma)$, where `$\mathcal{N}(\mu,\sigma)$' denotes the Guassian distribution with unknown mean $\mu$ and variance $\sigma$. A social scientist might then want to learn the value of $\mu$. The statistical error associated with each random variable $X_i$ is decomposed into a random and systematic component:

\begin{equation*}
    e_i = e_i^{random} + e^{\text{systematic}} = X_i - \mu
\end{equation*}

\noindent Given further knowledge of the particulars of this domain of social inquiry, the social scientist might impose additional assumptions about, for example, the shape of the distribution of the errors, or their covariance structure. These assumptions capture some of the flaws associated with a particular measurement instrument. Ultimately, a social scientist could make a post-hoc correction for measurement error by subtracting off (often called `de-biasing') the systematic component of the error:

\begin{equation*}
    X_i^{\text{new}} = X_i - e^{\text{systematic}}
\end{equation*}

\noindent Which will provide a more accurate estimate of $\mu$.
\end{example}

This example is paradigmatic of how measurement error is handled in the social sciences\footnote{In the words of political scientist Christopher Achen, ``measurement error is primarily a fault of the instruments, not of the respondents'' \autocite[p1229]{achen1975}.}. The post-hoc measurement error correction can be read as, effectively, claiming the underlying data generating process measured by the instrument is actually $X_i \sim \mathcal{N}(\mu - e^{\text{systematic}},\sigma)$ once the causal effect of the instrument on the data generating process is accounted for. However, this kind of de-biasing is only a minor change. The underlying model of the data generating process as a Guassian distribution is never challenged and employing an alternative mechanistic model of the data generating process is an entirely different class of proposition. The problem of reflexive measurement is a deeper one than the framework of measurement error allows. Ultimately, different instruments may radically alter the data generating process far beyond what measurement error is designed to capture. 

Furthermore, in example \ref{ICE} above, we saw how the survey instrument may be less important than the institution or person conducting the measurement. Consider the following alternative example,

\begin{example}[Free Legal Aid]\label{free}
A lawyer wishes to help undocumented immigrants naturalize in the US. They conduct a national phone survey to determine where to allocate their resources. They begin their survey by declaring "I am a lawyer offering free legal aid to undocumented immigrants to help them naturalize".
\end{example}

\noindent As in the ICE example (\ref{ICE}), the question asked on the survey could be letter-for-letter identical. The different preambles may result in entirely different pictures of the US at the level of data collection. Measurement error is not only narrowly concerned with instruments but also concerned with very specific (i.e., random, systematic) forms of error. That these two examples may have entirely different data generating processes cannot be understood within the existing framework of measurement error.

\section{Reflexive Measurement Instruments and Social 
Science}\label{sec_social_science}

In this section we focus on reflexive measurement \textit{instruments} and their role in social science. To do so, it is important to note that in the social sciences, measurement instruments are diverse: ``mathematical formulae, statistical formulae, and accounting rules, as well as the rules of data elicitation and manipulation'' \autocite[p237]{morgan2001}. Measurement instruments of social science are financial indices and economic indicators used by economists to measure the performance of an economy or market. They are the surveys administered by political scientists to gauge public opinion. They are the estimators chosen by statisticians to estimate statistical quantities of interest. And depending on the instrument used to measure the phenomena, the quality of data may vary (sometimes drastically).

As illustrated in section \ref{sec_problem} above, concerns of reflexive measurement arise even when considering the person or institution doing the measuring. However, the singular nature of an institution or person makes its consideration from a point of philosophical abstraction of limited utility. Surveys are ubiquitous, organizations which deport undocumented immigrants are not. Although this paper attempts to provide a single unified explanation of reflexivement measurement which illuminates both instruments and the entities that wield them, the focus of this section is squarely on the former.

I believe the solution to problems reflexive measurement lies in developing a scientific understanding of how our measurement instruments causally affect the underlying data generating process of the phenomena they are used to investigate. Measurement is central to building our understanding of the social world. Indeed, on one account of the importance of measurement instruments in economics, ``[m]easurement, maybe even more than theory, contributes to the making of the new ontic furniture for the economic world'' \autocite[p248]{morgan2001}. However, measurement without a social scientific model of how it affects the data generating process is, I argue, at best a stopgap solution to the problem of reflexive measurement.

Consider the following example of two survey questions asking about political attitudes towards the former US president George W. Bush. Firstly, a simple Likert scale for approval:

\begin{example}[Bush Approval 1]\label{ex_bush1}
What is your opinion of George W. Bush's presidency? \{Strongly Disapprove, Disapprove, Approve, Strongly Approve\}
\end{example}

\noindent Secondly, a similar Likert scale with a leading prompt:

\begin{example}[Bush Approval 2]\label{ex_bush2}
In 2003 George W. Bush invaded the sovereign nation of Iraq, resulting in the death of over half a million Iraqi civilians. What is your opinion of George W. Bush's presidency? \{Strongly Disapprove, Disapprove, Approve, Strongly Approve\}
\end{example}

\noindent The second question is an example of priming. Survey respondents are given information which may alter their beliefs or preferences prior to answering the survey question. These beliefs or preferences are the phenomena under investigation. Importantly, changes in beliefs or preferences which result from the application of a measurement instrument like a survey are not accounted for in classic philosophical accounts of systematic measurement error which are primarily concerned with the natural sciences \autocite[p858-9]{tal2019}. This is because once primed, a person cannot be primed in the same way again: they have learned new information about the world. The canonical natural science formulation of systematic error does not apply: in the social sciences instruments can never be used `in the same way, in the same case'.

From the perspective of the social sciences, one could attempt to subsume priming under concerns of (systematic) measurement error (see, for example, \autocite{groves2011}). In practice, it very often is. However, without an understanding of reflexive measurement we have no way of recognising that the underlying phenomena is altered by our measurements. Repeated application of the same measurement will fundamentally alter the phenomenon; there will no longer be any `error' to speak of. Theoretical models of rationality or learning can be used to explain this behaviour (i.e., Bayesian updating). What at first glance seems like a problem of measurement error is in fact a problem of reflexive measurement in disguise. 

A purely statistical approach---atheoretic in the social scientific sense---might advocate a de-biasing of question \ref{ex_bush1} and \ref{ex_bush2} separately (as outlined in the measurement error example \ref{ex_measure} above). In aggregate we might divine the net effect of priming the target population and subtract that component of systematic error off our estimate. However, since we lack a theoretical understanding of the relationship between our instrument and the phenomena and/or a data generating process, we do not have a theoretical reason to expect this de-biasing will hold in the future. Similarly, we cannot even generalize from question to question. In order to derive a prediction about how an instrument might perform in future we need a theory of how it interferes (or not) with a particular phenomenon and/or a data generating process.

A popular approach to dealing with measurement issues is to use \textit{robust} statistical methods. These methods are incredibly useful for social scientists worried about outliers or data that do not follow the data generating process assumed by the model. Thus, they might be considered helpful for social scientists worried about the affects of reflexive measurement on the data generating process. I believe, however, that a purely statistical approach to the problem of reflexive measurement fails to provide a social scientist with data that are ``reliable evidence'' (above) for the phenomenon under investigation. To fix an example of a popular robust statisitcal method, consider the approach of the Huber contamination model \autocite{huber1964}.

\begin{example}[Huber contamination model]\label{huber}
Let $X_1, \dots, X_n$ be independent random variables drawn from a distribution with cumulative distribution function given by 

\begin{equation*}
    F(t) = (1 - \epsilon)P(t) + \epsilon Q(t)
\end{equation*}

\noindent where $0 \leq \epsilon < 1$. $P$ is a known distribution function and $H$ is an unknown contaminating distribution.
\end{example}

Here, although most (i.e., $1-\epsilon)$ of the sample follows a known statistical distribution we explicitly model a small `contaminated' share $\epsilon$ to be drawn from an arbitrary distribution. Intuitively, one can think of this model as one where $\epsilon$ fraction of the sample are statistical outliers or anomalies. Huber's contamination model captures the following worry: ``[w]hat happens if the true distribution deviates slightly from the assumed [known] one?'' \autocite[p74]{huber1964} The sample mean is a classic (i.e., non-robust) statistical estimator which is notorious for exhibiting ``catastrophically'' \autocite[p74]{huber1964} poor performance in the presence of mild deviations. Thus, Huber proposes a class of statistical estimators known as $M$-estimators which are robust to large deviations. These methods offer the social scientist a kind of insurance against measurement concerns which might effect their data quality.

By way of contrast, consider an example where the measurement instrument is the choice of statistical estimator itself. If the estimator is publicly known then people can adapt their behavior in light of it. So it is clearly a reflexive measurement instrument. As we will see, we can explicitly model the relationship between the instrument and the data it generates with economic concept of \textit{incentive compatibility}\footnote{For a pedagogical introduction to the concept of incentive compatibility as it is used in mechanism design see, for example, \autocite{borgers2015}.}.

\begin{example}[Incentive-compatible estimation]\label{ic_estimators}
A statistician is trying to estimate the mean preferred temperature of occupants of a building. The statistician believes peoples' preferred temperatures follow a Guassian distribution with unknown mean $\mu$ and variance $\sigma$. A sample of $n$ occupants are randomly selected and asked their preferred temperature (i.e., $X_1, \dots, X_n \sim \mathcal{N}(\mu, \sigma)$).

However, each person sampled is told that the estimator the statistician will use for their estimate of the population mean is the sample mean. Notice that if you have a preference for, say, warmer temperatures, you are best off lying about your preferred temperature to raise the sample average. If your preferred temperature is, say 30C, it may be beneficial to you to lie and report an extreme temperature (e.g., 50C, 100C) to pull up the average of the sample.

Imagine the statistician instead uses the sample median as his estimate of the population mean and this is communicated to each person in the sample. Even if you have a preference for much warmer temperatures, you no longer gain by lying since the median is robust to large outlier values (see \autocite{caragiannis2016} for extended discussion of this result).
\end{example}

In this example the entire data generating process changes as a function of the estimator. The underlying phenomenon of interest (i.e., people's preferred temperature) remains unchanged. This is a concrete example of a social scientific model of a (statistical) measurement instrument's effect on the data generating process. The framework of incentive-compatible estimators and algorithms has been extended to explicitly causal settings \autocite{toulis2015}, forecasting problems \autocite{roughgarden2017}, and even bandit-type exploration algorithms \autocite{mansour2019}. It is a flexible approach to handling some types of reflexive measurement problems; however, it is assumed that people know the functional form of the statistical instrument (in this example, the mean or median). This is clearly a strong assumption and indicates these type of reflexive measurement concerns will not be encountered every day. Ultimately, the framework used in the example above yields a statistical estimator that satisfies the demands articulated earlier for an explicit model between the reflexive measurement instrument and the data generating process.

In example \ref{ic_estimators} above, there is an explicit model of the relationship between the statistical estimator and the data generating process in terms of benefits to people (i.e., their utility). The choice of statistical estimator (instrument) is recast as a game theoretic problem whereby the statistician and the people in the sample play a game. The statistician wants to estimate the population mean. People in the sample will only report their preferred temperature truthfully if they stand to benefit by it (or don't benefit by lying). Despite the game-theoretic formulation of the problem the goal of statistical inference remains the same. In the Huber contamination model introduced in example \ref{huber}, there is no scientific model of the effect of the instrument on the data generating process whatsoever. The model is entirely statistical and aims to mitigate `contamination' effects, broadly construed.

The generality of the Huber model may be a virtue in many settings. In contrast to the incentive-compatible estimator in example \ref{ic_estimators}, no assumptions needs to be made about the estimator's effect on the data generating process. In the natural sciences this may be entirely appropriate and in the social sciences there are many legitimate uses for robust statistics. Indeed, it is important not mistake this as an argument against the use of robust statistics in the social sciences. Often these may outperform classical (i.e., non-robust) methods. But they are not a substitute for building a model of the relationship between a reflexive measurement instrument and a data generating process it alters. At best they are a stopgap. Ultimately, I believe the fact that robust statistical methods perform well in many social science settings is an indication of the paucity of our scientific knowledge of the social world.

Building theoretical model of reflexive measurement instruments and their relationship to data generating processes can be done in any number of ways. In example \ref{ic_estimators} the theorized causal effect of the estimator was constructed in game theoretic terms, but it could have instead by a partial differential equation or a multi-agent simulation. In my view, the critically important consideration is that our understanding of the strengths and limitations of our instruments is framed in social scientific terms and theories. In the presence of problems of reflexive measurement `better' statistics will not make for better science absent models of how reflexive measurement instruments causally affect data generating processes.

\section{Conclusion}\label{sec_conc}

I have argued that in the social sciences acts of measurement causally affect the phenomenon under investigation and data generating processes associated with it. This is problem of reflexive measurement. Contemporary statistical approaches, often informed by concerns of measurement error, are insufficient to capture these causal relationships in their entirety. Thus, social scientists need a scientific understanding of the relationship between their instruments and the phenomena they are designed to measure. Ultimately, purely statistical corrections fail to create scientific knowledge in the face of problems of reflexive measurement.

Although the specific argument advanced here for the development of social scientific theories of reflexive measurement instruments is novel, the recognition that social scientists need better knowledge of their tools is not. Writing in 1975, political scientist Christopher Achen, acknowledged that 

\begin{quote}
``[m]ajor improvements in our understanding of political thinking may therefore come to depend upon a considerably more advanced theoretical knowledge of our measuring instruments than we have yet mustered.'' \autocite[p1231]{achen1975}
\end{quote}

\noindent Intuitively, we recognise that the New York sociologist in example \ref{NYC} is not getting reliable evidence of attitudes towards New York City. The survey question which primes respondents with information about George W. Bush's presidency in example \ref{ex_bush2} affects the very phenomenon it is designed to measure. Examples like these are ubiquitous in the social sciences. Their pervasiveness merits philosophical concern.

This essay is primarily focused on the use of statistics in the social sciences. Yet the problem of reflexive measurement affects qualitative social science research too. Qualitative researchers are painfully aware of this (see, for example, \autocite{king2021}.) I believe the challenge of developing a theory of reflexive measurement in qualitative research is significantly harder. This is for similar reasons to why reflexive measurement issues stemming from persons or institutions was neglected in this essay: institutions and persons are singular entities which render generalization and abstraction harder. Unfortunately, I do not have space to defend this view. Ultimately, exploring the problem of reflexive measurement in qualitative research remains an open problem for future research.


It is commonly said that `only a poor worker blames their tools'. This essay has argued that the toolkit of modern social science fails in the presence of a particular, pervasive type of measurement concern. Thankfully, there is reason to be optimistic. The methodological innovations outlined in section \ref{sec_social_science} are examples of instruments which are explicitly designed to mitigate these concerns \autocite{caragiannis2016, toulis2015, mansour2019}. This line of research is still in its early stages but it represents a significant step forward in taking seriously the problem of reflexive measurement.

\section{Acknowledgements}  

Dejan Makovec, Kevin Zollman, University of Pittsburgh History and Philosophy of Science Department (HPS) graduate students, Bele Wollesen, Marina Dubova, Liam Kofi Bright, Antonio Alexis Mahana, Agne Sabaliauskaite.

\section{References} 

\printbibliography[heading=none]

\end{document}